\documentclass[11pt,a4paper]{emulateapj}
\slugcomment{{\sc Accepted to The Astrophysical Journal:} February 6, 2013}

\usepackage{natbib}

\newcommand{\teff}{${T}_{\mathrm{eff}}$}
\newcommand{\logg}{$\log{g}$}
\newcommand{\zzc}{ZZ~Ceti}
\newcommand{\msun}{${M}_{\odot}$}
\newcommand{\mstar}{${M}_{\star}$}

\newcommand{\omc}{($O-C$)}
\newcommand{\wdf}{SDSS~J011100.63+001807.2}
\newcommand{\wdo}{WD~0111+0018}
\newcommand{\fa}{$f_1$}
\newcommand{\fb}{$f_2$}
\newcommand{\fy}{$2f_1$}
\newcommand{\fz}{$f_1+f_2$}
\newcommand{\bruntv}{Brunt-V\"{a}is\"{a}l\"{a}}

\shorttitle{Rapid period change in the DAV WD~0111+0018}
\shortauthors{Hermes et al.}

\begin{document}

\title{A NEW TIMESCALE FOR PERIOD CHANGE IN THE PULSATING DA WHITE DWARF WD~0111+0018}
\author{J. J. Hermes\altaffilmark{1,2}, M. H. Montgomery\altaffilmark{1,2}, Fergal Mullally\altaffilmark{3,4}, D. E. Winget\altaffilmark{1,2}, and A. Bischoff-Kim\altaffilmark{5}}

\altaffiltext{1}{Department of Astronomy, University of Texas at Austin, Austin, TX\,-\,78712, USA}
\altaffiltext{2}{McDonald Observatory, Fort Davis, TX\,-\,79734, USA}
\altaffiltext{3}{SETI Institute, 189 Bernardo Avenue, Suite 100, Mountain View, CA 94043, USA}
\altaffiltext{4}{NASA Ames Research Center, Moffett Field, CA 94035, USA}
\altaffiltext{5}{Chemistry, Physics and Astronomy Department, Georgia College \& State University, Milledgeville, GA, 31061, USA}

\email{jjhermes@astro.as.utexas.edu}


\begin{abstract}

We report the most rapid rate of period change measured to date for a pulsating DA (hydrogen atmosphere) white dwarf (WD), observed in the 292.9 s mode of WD~0111+0018. The observed period change, faster than 10$^{-12}$ s s$^{-1}$, exceeds by more than two orders of magnitude the expected rate from cooling alone for this class of slow and simply evolving pulsating WDs. This result indicates the presence of an additional timescale for period evolution in these pulsating objects. We also measure the rates of period change of nonlinear combination frequencies and show that they share the evolutionary characteristics of their parent modes, confirming that these combination frequencies are not independent modes but rather artifacts of some nonlinear distortion in the outer layers of the star.

\end{abstract}



\section{Introduction}

White dwarf (WD) stars represent the final evolutionary stage of all single low-mass stars, and are thus representative of the future of the majority of stars in our Galaxy. They are primarily composed of the inert by-products of hydrogen and helium fusion, and their evolution is dictated by the rate at which these carbon and oxygen ions lose their residual thermal energy \citep{Mestel52}. This cooling takes billions of years.

However, we can witness this cooling on human timescales by watching the WDs that pulsate. These variable stars come in several classes, based on their dominant atmospheric composition (their strong surface gravity and rapid gravitational settling leads to chemically homogenous photospheres). Here we will restrict our discussion to the coolest class of such pulsators: the hydrogen-atmosphere DAV (or \zzc) stars, found in an instability strip between $10800-12300$ K for standard \logg\ = 8.0 \citep{Koester01,Bergeron04,Mukadam04,Gianninas05,Gianninas06,Mukadam06}.

Aside from their variability, which is brought on by a hydrogen partial-ionization zone in the star's non-degenerate atmosphere, DAVs appear to be normal WDs. They are therefore believed to be representative of evolution for all DA WDs \citep{Robinson79,Fontaine85,Bergeron04}. Cooling from neutrino emission is expected to be negligible for DA WDs within the \zzc\ instability strip \citep{Winget04}, and their evolution should be dominated by radiative surface emission.

The C/O-core DAVs undergo multi-periodic, non-radial $g$-mode pulsations with periods between roughly $100-1400$ s. Seismology using these $g$-mode pulsations has enabled us to constrain the mass, core and envelope composition, rotation rate, and the behavior of convection in these objects (see reviews by \citealt{WinKep08}, \citealt{FontBrass08} and \citealt{Althaus10}). These modes are often stable in period and amplitude, especially for the hot DAVs (hDAVs) near the blue-edge of the \zzc\ instability strip. hDAVs are expected to show an extremely slow period drift caused by the gradual cooling of the star.

Two of the longest-studied hDAVs are G117-B15A and \zzc\ itself (also referred to as R548), which have been observed since the early 1970s. Both show considerable stability in their largest-amplitude modes. The rate of change of period with time, $dP/dt$, of the dominant 215.2 s mode in G117-B15A has been measured to be $(4.19\pm0.73) \times 10^{-15}$ s s$^{-1}$ \citep{Kepler11}. R548 has been observed since 1970, but more sparingly: For the 213.1 s periodicity in R548, \citet{Mukadam12} determine a $dP/dt$ = ($3.3 \pm 1.1) \times 10^{-15}$ s s$^{-1}$. These values have taken decades of observations, and are in line with theoretical predictions of rates of period change from cooling alone in DAVs of $dP/dt$ $< 10^{-14}$ s s$^{-1}$ \citep{Bradley92,Kim08,Corsico12}.

After more than nine years of monitoring, we have found evidence for an hDAV that has a rate of period change inconsistent with cooling alone. The period change in the highest-amplitude periodicity for that star, \wdf\ (hereafter \wdo; $g$\,=\,18.7 mag), exceeds $10^{-12}$ s s$^{-1}$. This observed rate of change of period with time is several orders of magnitude faster than predicted from cooling alone for this 11810 K WD. It therefore signifies some physical process operating on a considerably shorter timescale.

In this paper we present our observations of period evolution in the four highest-amplitude periodicities present in \wdo. In Sections 2 and 3 we outline our observations and analysis, and we reserve Sections 4 and 5 for a discussion of possible explanations for this rapid period change and our conclusions.



\section{Time-Series Photometric Observations}

\citet{Mukadam04} discovered pulsations in \wdo, resulting from a search for variable DAs. The WD was initially targeted as a result of model atmosphere fits to spectra from the Sloan Digital Sky Survey: with \teff\ = $11510\pm110$ K and \logg\ = $8.26\pm0.06$, the object was predicted and confirmed to be a DAV within the \zzc\ instability strip.

Using new treatments of line broadening theory, \citet{Tremblay11} refined the temperature and surface gravity of \wdo\ to \teff\ = $11810\pm190$ K and \logg\ = $8.17\pm0.07$, which corresponds to a mass of $M_\mathrm{WD} = 0.71\pm0.04$ \msun\ and a cooling age of 510 Myr.

Including the initial discovery observations in early 2003, the White Dwarf Group at the University of Texas at Austin has logged a total of more than 136 hr of observations of \wdo\ on 39 separate nights. As the journal of observations in Table~\ref{tab:jour} indicates, the data span more than nine years. One motivation for these observations was to use the stable pulsation modes of hDAVs as precise clocks, searching for evidence of planetary companions by looking for periodic variations in the pulse arrival times \citep{Winget03,Mullally08}. The project has also effectively opened a window on new evolutionary timescales for more than a dozen hDAVs, including \wdo.

\begin{deluxetable}{lrccc}
\tabletypesize{\scriptsize}
\tablecolumns{5}
\tablewidth{0.38\textwidth}
\tablecaption{Journal of observations. \label{tab:jour}}
\tablehead{
\colhead{Subgroup} & \colhead{UT Date} & \colhead{Length} & \colhead{Seeing} & \colhead{Exp.}
\\ \colhead{} & \colhead{} & \colhead{(hr)} & \colhead{(\arcsec)} & \colhead{(s)} }
\startdata
2003a 	& 27 Jan 2003 & 1.5 & 2.3 & 15 \\
 		& 02 Feb 2003 & 0.9 & 1.6 & 20 \\
2004a 	& 20 Nov 2003 & 2.9 & 2.3 & 15 \\
 		& 30 Nov 2003 & 3.0 & 2.4 & 15 \\
2005a 	& 12 Dec 2004 & 5.0 & 2.5 & 15 \\
 		& 14 Dec 2004 & 4.2 & 2.7 & 15 \\
 		& 16 Dec 2004 & 3.4 & 1.4 & 15 \\
2006a 	& 03 Dec 2005 & 3.5 & 1.8 & 15 \\
 		& 06 Dec 2005 & 5.4 & 1.5 & 15 \\
2006b 	& 30 Dec 2005 & 3.4 & 2.0 & 15 \\
 		& 31 Dec 2005 & 2.8 & 2.0 & 15 \\
 		& 03 Jan 2006 & 2.6 & 1.6 & 15 \\
2007a 	& 19 Sep 2006 & 3.1 & 1.5 & 15 \\
 		& 22 Sep 2006 & 2.7 & 1.7 & 15 \\
2007b 	& 25 Nov 2006 & 6.1 & 1.3 & 10 \\
2010a 	& 23 Aug 2009 & 2.2 & 1.6 & 15 \\
 		& 24 Aug 2009 & 4.2 & 1.4 & 15 \\
2010b 	& 15 Sep 2009 & 5.5 & 1.5 & 15 \\
2010c 	& 11 Nov 2009 & 3.5 & 1.6 & 15 \\
 		& 17 Nov 2009 & 3.7 & 2.7 & 30 \\
 		& 21 Nov 2009 & 7.4 & 1.6 & 30 \\
2010d 	& 12 Jan 2010 & 3.3 & 2.1 & 15 \\
 		& 13 Jan 2010 & 3.6 & 1.8 & 15 \\
2011a 	& 12 Sep 2010 & 3.6 & 1.1 & 15 \\
 		& 13 Sep 2010 & 6.0 & 1.6 & 15 \\
2011b 	& 07 Oct 2010 & 4.5 & 1.3 & 15 \\
 		& 08 Oct 2010 & 4.3 & 1.6 & 15 \\
 		& 09 Oct 2010 & 5.3 & 1.3 & 15 \\
2011c 	& 10 Nov 2010 & 1.7 & 1.8 & 15 \\
 		& 11 Nov 2010 & 3.1 & 1.3 & 15 \\
2011d 	& 02 Jan 2011 & 2.8 & 3.0 & 20 \\
 		& 04 Jan 2011 & 4.1 & 1.6 & 15 \\
 		& 05 Jan 2011 & 1.7 & 1.7 & 20 \\
2012a 	& 02 Oct 2011 & 3.8 & 1.2 & 15 \\
 		& 03 Oct 2011 & 2.6 & 1.2 & 15 \\
2012b 	& 23 Oct 2011 & 4.5 & 1.5 & 15 \\
2012c 	& 01 Feb 2012 & 1.2 & 1.4 & 15 \\
 		& 02 Feb 2012 & 1.2 & 2.5 & 15 \\
 		& 03 Feb 2012 & 1.8 & 2.1 & 15
\enddata
\end{deluxetable}

All of our data on \wdo, more than 33,100 images, have been taken using the same instrument (Argos, a frame-transfer CCD; see \citealt{Nather04}) with the same filter (a 2 mm \textsl{BG40}, to reduce sky noise) on the same telescope (the 2.1m Otto Struve at McDonald Observatory), allowing for an especially coherent data set. Depending on the conditions, we have used $10-30$ s exposures, as indicated in Table~\ref{tab:jour}.

The raw science frames are calibrated by dark subtraction and flat-fielding. We perform weighted, circular, aperture photometry on the calibrated frames using the external IRAF package $\textit{ccd\_hsp}$ written by Antonio Kanaan (the reduction method is outlined in \citealt{Kanaan02} and \citealt{Mullally05}). We divided the sky-subtracted light curves using two bright comparison stars in the field, SDSS~J011055.57+001850.6 ($g=15.0$ mag) and SDSS~J011056.17+001955.8 ($g=16.2$ mag), to correct for transparency variations. Using the WQED software suite \citep{Thompson09}, we fit and subtract out a low-order polynomial (at the timescale of several hours) to each run to remove any long-term trend caused by atmospheric extinction, and apply a timing correction to each observation to account for the motion of the Earth around the barycenter of the solar system \citep{Stumpff80}. 



\section{Light Curve Analysis}

\subsection{The Pulsation Spectrum of \wdo}
\label{sec:FT}

\begin{figure}[t]
\centering{\includegraphics[width=\columnwidth]{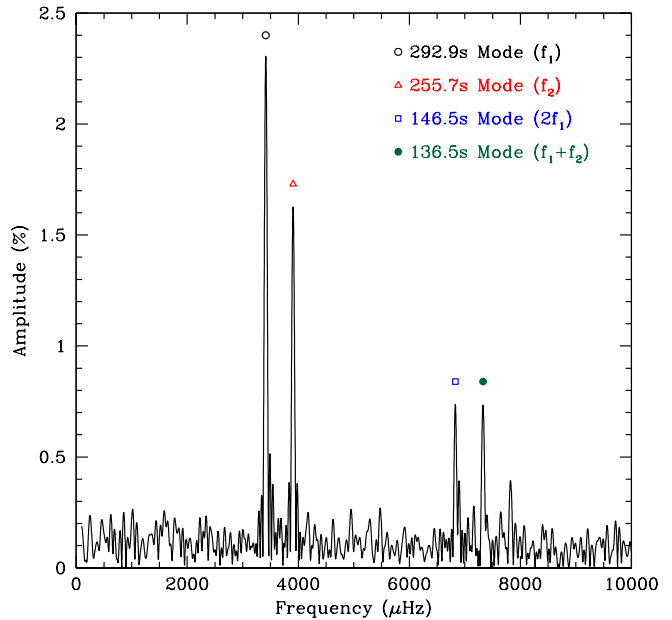}}
\caption{The pulsation spectrum of \wdo\ for a typical run, this on 15 Sep 2009. A Fourier transform of this 5.5-hr run illustrates that the four periodicities of interest, seen marked, can be resolved in a single night. \label{fig:FTrun}}
\end{figure}

The light curve of \wdo\ features modulation at four distinct periodicities between $100-300$ s, which can be well-resolved in a 3-hr run in good conditions (see Figure~\ref{fig:FTrun}). In order to most accurately identify these four periods, we begin by computing a Fourier transform of our entire data set. Using these results as an initial guess, we then perform a nonlinear least-squares fit for the frequency, amplitude, and phase of the four periodicities present.

This allows us to isolate an initial estimate of the periods of the four highest-amplitude pulsations present in \wdo, which we use to construct an initial \omc\ diagram. The amplitudes for these four periodicities never fall below 0.5\%; the amplitudes listed in Table~\ref{tab:freq} are weighted-mean amplitudes using each subgroup.

\begin{deluxetable*}{lcccc}
\tablecolumns{5}
\tabletypesize{\scriptsize}
\tablewidth{0.65\textwidth}
\tablecaption{Mode periods and observed rates of period change in \wdo \label{tab:freq}}
\tablehead{
\colhead{Mode} & \colhead{$t_0$} & \colhead{$P_0$} & \colhead{Amplitude} & \colhead{$dP/dt$} \\ \colhead{} & \colhead{(BJD$_{\rm{TDB}}$)} & \colhead{(s)} & \colhead{(\%)} & \colhead{($10^{-12}$ s s$^{-1}$)} }
\startdata
\fa & 2452666.562825(21) & 292.944305(9) & 2.573 & 4.280(42) \\
\fb & 2452666.562314(31) & 255.663971(2) & 1.428 & 0.340(58) \\
\fy & 2452666.562701(42) & 146.472153(3) & 0.633 & 2.075(44) \\
\fz & 2452666.562624(29) & 136.518707(5) & 0.730 & 1.109(30)
\enddata
\end{deluxetable*}

We perform this analysis independently for all 18 subgroups of our data. The subgroups are created by combining data from as many contiguous nights as possible; no combined subgroup spans more than two weeks. These subgroups have been identified in Table~\ref{tab:jour}. In each case, the periods we determine for each subgroup match the periods we have found for the entire data set, within the uncertainties.

We fix these periods and compute a simultaneous linear least-squares fit for each subgroup to create an \omc\ diagram. Thus, the uncertainties in our \omc\ diagrams represent formal uncertainties on the least-squares fit for the phase on at least one---and as many as three---nights of data, separated by as many as 10 days.

We will refer to the two highest-amplitude periodicities as the parent modes \fa\ and \fb, and the two smaller-amplitude periodicities \fy\ and \fz\ as the combination frequencies, such that the frequency for \fy\ is (within the uncertainty) exactly twice the frequency of \fa\ and the frequency for \fz\ is the sum of the frequencies of \fa\ and \fb. We will discuss in more detail these combination frequencies and how they arise in Section~\ref{sec:combs}.

Additionally, we note the presence of an additional combination frequency $2f_2$ at 127.8 s, visible but not labeled in Figure~\ref{fig:FTrun}. In fact, we detect at least five additional nonlinear combination frequencies. Using all of our light curves from 2009 to early 2012 to detect as many combination frequencies as possible, we find: $f_1=2.739(29)$\%, $f_2=1.526(31)$\%, $2f_1=0.472(27)$\%, $f_1+f_2=0.732(28)$\%, $2f_2=0.193(25)$\%, $3f_1=0.143(44)$\%, $2f_1+f_2=0.295(98)$\%, $f_1+2f_2=0.147(44)$\%, $3f_2=0.068(28)$\%. However, the amplitudes of most of the combination frequencies are too low to produce an \omc\ diagram for all subgroups, so we do not include them in the \omc\ analysis.

\subsection{Constructing an \omc\ Diagram}

\begin{figure*}[t]
\centering{\includegraphics[width=\textwidth]{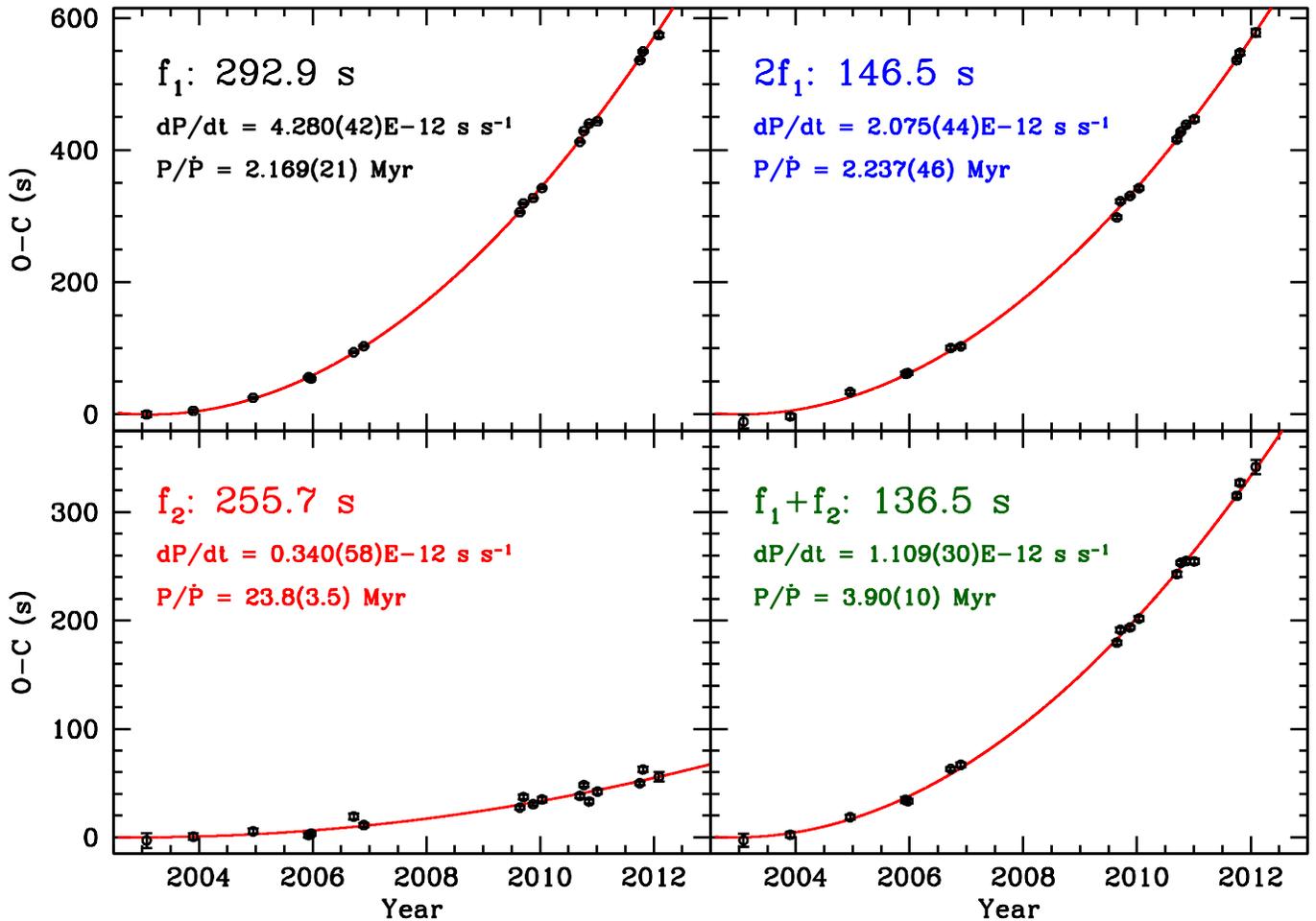}}
\caption{\omc\ diagrams for the four highest-amplitude periodicities present in \wdo. The \fa\ and \fy\ \omc\ diagrams are nearly identical, which strongly suggests that nonlinear combination frequencies in DAVs are not independent pulsation modes but are directly tied to their parent modes. A best-fit parabola yields a rate of change of period with time: See Table~\ref{tab:freq} for full solutions. \label{fig:omc}}
\end{figure*}

We demonstrate a secular change in the pulsation periods of \wdo\ by constructing \omc\ diagrams, where we compare the observed time of maximum for a pulsation ($O$) to when we expect such a maximum assuming that the pulsation obeys a constant period ($C$).

Following \citet{Kepler91}, if the pulsation period is changing slowly with time, we can expand the observed time of maximum of the E$^{th}$ pulse, $t_E$, in a Taylor series around $E_0$:
\begin{eqnarray}
t_{max} \mid _{E} \; = \; t_{max} \mid _{E_0} + \; \frac{dt_{max}}{dE} \mid _{E_0} (E - E_0) + \nonumber \\ \frac{1}{2} \frac{d^2t_{max}}{dE^2} \mid _{E_0} (E - E_0)^2 + \; ...
\end{eqnarray}
where the epoch is $E=t/P$ and the change in arrival time with epoch, $dt/dE$, is the period, $P$. If we drop all terms higher than second order (assuming that $\ddot{P}$ is negligible), we arrive at the classic \omc\ equation
\begin{equation}
O \; - \; C = \Delta t_{0} + \, \Delta P_0 \, E + \, \frac{1}{2} P_0 \dot{P} E^2
\end{equation}
where $t_0$ is the time of first maximum, $\Delta t_{0}$ is the uncertainty in this time, $P_0$ is the pulsation period at this time of first maximum and $\Delta P_0$ is the error in the observed period. Thus, any secular change in the period, $dP/dt$, will cause a parabolic curvature in an \omc\ diagram.

We construct an initial \omc\ diagram using the periods identified in Section~\ref{sec:FT}. We then iteratively adjust $t_0$ and $P_0$ by the zeroth- and first-order terms from our best second-order fit until the adjustments are smaller than the uncertainty in these terms; these uncertainties result from the covariance matrix. Our recomputed, final \omc\ diagrams use these final ephemera and periods, which can be found for each mode in Table~\ref{tab:freq}. The final \omc\ diagrams are shown in Figure~\ref{fig:omc}. Table~\ref{tab:omc} presents the times of maximum from each subset of our observations.

\begin{deluxetable}{lcccc}
\tablecolumns{5}
\tabletypesize{\scriptsize}
\tablewidth{0.42\textwidth}
\tablecaption{Observed times of maximum in \wdo \label{tab:omc}}
\tablehead{
\colhead{} & \colhead{Time of Maximum} & \colhead{Epoch} & \colhead{\omc} & \colhead{$\sigma$} \\ \colhead{} & \colhead{(BJD$_{\rm{TDB}}$)} & \colhead{} & \colhead{(s)} & \colhead{(s)} }
\startdata
\fa & 2452668.431022	&	551	&	-0.11	&	4.41	\\
 & 2452968.919383	&	89176	&	5.29	&	1.96	\\
 & 2453353.385279	&	202569	&	25.11	&	1.62	\\
 & 2453709.441810	&	307583	&	56.14	&	1.30	\\
 & 2453725.336726	&	312271	&	53.99	&	1.34	\\
 & 2453999.274020	&	393065	&	94.00	&	1.38	\\
 & 2454064.667839	&	412352	&	103.15	&	1.33	\\
 & 2455067.170353	&	708026	&	305.96	&	1.09	\\
 & 2455089.880472	&	714724	&	319.29	&	1.69	\\
 & 2455152.022733	&	733052	&	327.43	&	0.87	\\
 & 2455209.164001	&	749905	&	342.55	&	1.31	\\
 & 2455452.386565	&	821640	&	412.39	&	1.24	\\
 & 2455477.944792	&	829178	&	429.02	&	1.00	\\
 & 2455511.887808	&	839189	&	440.13	&	1.29	\\
 & 2455565.211171	&	854916	&	443.63	&	1.37	\\
 & 2455837.226626	&	935143	&	536.16	&	1.07	\\
 & 2455857.749832	&	941196	&	549.35	&	1.67	\\
 & 2455959.724574	&	971272	&	574.11	&	2.94	\\
\fb & 2452668.432414	&	632	&	-2.94	&	6.75	\\
 & 2452968.917516	&	102179	&	0.57	&	3.07	\\
 & 2453353.384106	&	232107	&	5.55	&	2.53	\\
 & 2453709.440530	&	352434	&	1.93	&	2.15	\\
 & 2453725.333732	&	357805	&	3.46	&	2.00	\\
 & 2453999.273126	&	450381	&	19.25	&	2.58	\\
 & 2454064.668563	&	472481	&	11.27	&	2.06	\\
 & 2455067.167424	&	811269	&	27.44	&	1.75	\\
 & 2455089.875468	&	818943	&	37.12	&	2.11	\\
 & 2455152.018898	&	839944	&	30.44	&	1.34	\\
 & 2455209.158663	&	859254	&	34.84	&	2.40	\\
 & 2455452.382727	&	941450	&	38.19	&	1.96	\\
 & 2455477.937402	&	950086	&	48.08	&	1.66	\\
 & 2455511.880763	&	961557	&	33.06	&	1.86	\\
 & 2455565.206336	&	979578	&	42.18	&	2.23	\\
 & 2455837.222237	&	1071504	&	49.81	&	1.65	\\
 & 2455857.743562	&	1078439	&	62.61	&	2.60	\\
 & 2455959.716124	&	1112900	&	55.88	&	4.21	\\
\fy & 2452668.432466	&	1103	&	-11.08	&	10.53	\\
 & 2452968.917468	&	178351	&	-3.06	&	3.98	\\
 & 2453353.385256	&	405138	&	33.60	&	3.11	\\
 & 2453709.441750	&	615166	&	61.32	&	2.54	\\
 & 2453725.336700	&	624542	&	62.18	&	2.86	\\
 & 2453999.273975	&	786130	&	100.45	&	2.92	\\
 & 2454064.669411	&	824705	&	102.78	&	2.50	\\
 & 2455067.170145	&	1416052	&	297.94	&	2.62	\\
 & 2455089.878699	&	1429447	&	322.50	&	2.64	\\
 & 2455152.022654	&	1466104	&	330.52	&	1.89	\\
 & 2455209.163883	&	1499810	&	342.33	&	2.46	\\
 & 2455452.386495	&	1643280	&	416.25	&	2.40	\\
 & 2455477.942967	&	1658355	&	427.70	&	1.97	\\
 & 2455511.885977	&	1678377	&	438.32	&	2.51	\\
 & 2455565.211093	&	1709832	&	446.75	&	3.74	\\
 & 2455837.228208	&	1870287	&	536.15	&	2.18	\\
 & 2455857.747999	&	1882391	&	547.15	&	3.50	\\
 & 2455959.722809	&	1942543	&	577.84	&	5.71	\\
\fz & 2452668.431822	&	1183	&	-2.87	&	5.98	\\
 & 2452968.918406	&	191355	&	2.36	&	2.44	\\
 & 2453353.384661	&	434676	&	18.53	&	2.29	\\
 & 2453709.441119	&	660017	&	34.57	&	2.73	\\
 & 2453725.335107	&	670076	&	33.47	&	2.10	\\
 & 2453999.273507	&	843446	&	62.98	&	1.87	\\
 & 2454064.668225	&	884833	&	66.87	&	2.13	\\
 & 2455067.168742	&	1519295	&	179.62	&	2.06	\\
 & 2455089.877753	&	1533667	&	191.35	&	2.44	\\
 & 2455152.020653	&	1572996	&	193.66	&	1.54	\\
 & 2455209.162674	&	1609160	&	201.74	&	2.14	\\
 & 2455452.384498	&	1763090	&	242.81	&	2.44	\\
 & 2455477.940796	&	1779264	&	253.35	&	1.72	\\
 & 2455511.884039	&	1800746	&	254.73	&	1.85	\\
 & 2455565.210077	&	1834495	&	254.55	&	2.49	\\
 & 2455837.225881	&	2006648	&	315.02	&	1.77	\\
 & 2455857.746487	&	2019635	&	326.94	&	2.50	\\
 & 2455959.720129	&	2084172	&	341.82	&	6.59	
\enddata

\end{deluxetable}

\subsection{Observed Rates of Period Change}
\label{sec:rates}

\citet{Mullally08} published an \omc\ diagram of \fa\ in \wdo, noting the large curvature but urging caution given the sparse coverage. Still, a rather high rate of change of period with time was observed: $(3.87\pm 0.43) \times 10^{-12}$ s s$^{-1}$.

We have added another 86 hr of observations, more than doubling the coverage, and confirm this large trend in the \omc\ diagram of \fa. We have additionally been able to make \omc\ diagrams for three other periodicities. These diagrams can be found in Figure~\ref{fig:omc}.

Using Equation 2, a best-fit parabola to each \omc\ diagram yields a rate of change of period with time. We list these results in Table~\ref{tab:freq}. Our results for \fa\ actually fall within the errors of the \citet{Mullally08} result, at $(4.280\pm 0.042) \times 10^{-12}$ s s$^{-1}$ (or $0.135\pm0.001$ ms yr$^{-1}$). We also strongly establish the need for a second-order fit rather than a first-order fit to the 18 data points: The best second-order fit for the \omc\ diagram of \fa\ has $\chi^2=145$ (15 degrees of freedom, or d.o.f.), whereas the best linear fit has $\chi^2=10292$ (16 d.o.f.).

Although the \omc\ diagrams for \wdo\ appear parabolic, as we would expect if the pulsations periods were changing secularly, we would also like to establish the timescale of variability if these observed trends are instead periodic in nature, since a portion of a sinusoid can mimic a parabola. For \fa\ we find that the data can be well fit by a sinusoid with a 101 year period and a 3931 s \omc\ amplitude, with a $\chi^2=142$ (15 d.o.f.). However, we will only continue our analysis in the framework of a secular period change.

In the future, a third-order fit including the acceleration in the period change may be needed to best represent the data. This would enter the \omc\ equation as 
$$ O \; - \; C = \Delta t_{0} + \, \Delta P_0 \, E + \, \frac{1}{2} P_0 \dot{P} E^2 + \, \frac{1}{6} ( \ddot{P} P_0^2 + \dot{P}^2 P_0) E^3$$
Currently, a third-order polynomial fit to the \omc\ diagram of \fa\ has $\chi^2=143$ (14 d.o.f.), which reduces to a value still slightly higher than that for a second-order fit. The best third-order fit would correspond to $\ddot{P} = (-2.3\pm 1.4) \times 10^{-21}$ s s$^{-2}$, with a second-order term corresponding to $dP/dt = (4.65\pm 0.23) \times 10^{-12}$ s s$^{-1}$.

We have also attempted a direct measurement of $dP/dt$ by computing a nonlinear least-squares fit to find the period for each subgroup. A best-fit line to these periods over time yields a rate of change of period with time. Unfortunately, there is simply not enough data for this direct measurement. The best result comes for \fa, where we determine $dP/dt = (5.5\pm6.8) \times 10^{-12}$ s s$^{-1}$, consistent with the \omc\ value but not yet significant.

The other independent parent mode in \wdo, \fb, shows a much slower period change than \fa. Still, it is significantly faster than predicted by cooling theory, with a best-fit parabola that corresponds to $(3.40\pm 0.58) \times 10^{-13}$ s s$^{-1}$ (or $0.011\pm0.002$ ms yr$^{-1}$). In this case, a second-order fit is only marginally better than a linear fit: The best-fit parabola has $\chi^2=93.8$ (15 d.o.f.), whereas the best-fit straight line has $\chi^2=128.7$ (16 d.o.f.).

Noise can cause scatter about a straight line in an \omc\ diagram, especially over a relatively short baseline (e.g., \citealt{Kepler91}). Since a bad period can cause a dominant linear term, we caution that the rate of change of the \fb\ periodicity may still be somewhat slower than the value cited in Table~\ref{tab:freq} if the period we use for our \omc\ diagrams is off by as little as 0.06 ms. Therefore, we recommend that our value be construed as an upper limit on the rate of change of period with time of \fb. As with the other parent mode, we also investigated the best periodic rather than secular fit, and found a sinusoid with a 9540 year period and 3051490 s \omc\ amplitude, with a $\chi^2=93.8$ (15 d.o.f.). (If such a trend were from an external companion, it would require a 10.8 \msun\ unseen companion at 401 AU.)

A Monte Carlo simulation of the uncertainties using the software package Period04 \citep{Lenz05} indicates that our formal least-squares uncertainties on the phases used to construct the \omc\ diagrams are underestimated by at most 10\%. Thus, the scatter about our residuals (evident by the large $\chi^2_{\mathrm{red}}$ quoted above for the second-order fits, e.g. $\chi^2_{\mathrm{red}}=9.7$ for \fa) may not be entirely the result of underestimated uncertainties. The deviation about the residuals is consistent with the observed scatter in other long-studied DAVs, such as G117-B15A (see Figure 2 of \citealt{Kepler05}).

\begin{figure}[t]
\centering{\includegraphics[width=\columnwidth]{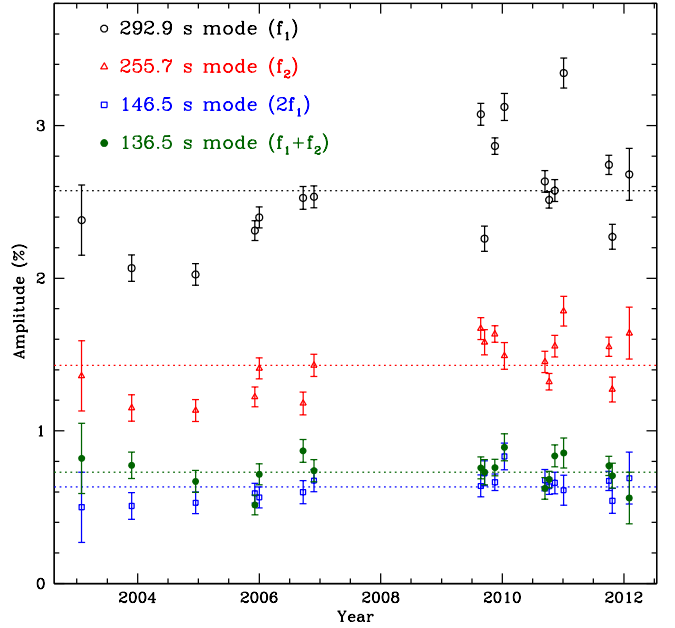}}
\caption{The amplitude evolution of \wdo\ over nine years of observations. The dotted lines represent the weighted mean amplitude over the entire data set, shown in Table~\ref{tab:freq}. While our data show that the amplitudes are not stable over the entire data set, they rule out a large-scale amplitude increase as the cause of the large trends in the \omc\ diagrams shown in Figure~\ref{fig:omc}. \label{fig:amps}}
\end{figure}

We have investigated the stability of the four highest-amplitude periodicities in \wdo\ by plotting the amplitudes of each determined by a linear least-squares fit to each subgroup, using the periods in Table~\ref{tab:freq}. This result is shown in Figure~\ref{fig:amps}. It is meaningful since all observations were obtained and reduced in an identical manner. We have also marked the weighted mean amplitude for each mode with dotted lines.

There are considerable deviations about this mean compared to the formal uncertainties, such that only the periodicity at \fy\ has a constant amplitude in a statistically meaningful way. While the amplitudes of the \fa\ and \fb\ modes do not appear constant over our entire data set, the deviations from the mean do not appear coherent, and cannot explain the large-scale parabolic changes we observe in observed times-of-maxima.

Finally, in line with the original intentions of this data set, we have investigated whether periodic changes in the times of maxima can explain the scatter observed about the residuals after subtracting the dominant parabolic terms. This scatter might reveal the presence of an unseen companion. However, a Fourier transform of all four \omc\ residuals, after subtracting the dominant parabolic terms identified in Figure~\ref{fig:omc}, show no peaks above twice the mean noise level of 1.5 s. We can thus conservatively rule out the presence of any companion more massive than Jupiter at a current orbit between $2-10$ AU; the scatter is predominantly incoherent and cannot currently be attributed to an unseen companion.



\section{Discussion}

\subsection{Nonlinear Combination Frequencies}
\label{sec:combs}

Before focusing on the rapid rates of period change, we first draw attention to the observation that $dP/dt$ for the \fy\ and \fz\ combination frequencies evolve exactly in lockstep with their parent modes. For example, we expect the \fy\ combination frequency to have half the rate of change of period with time of its parent mode, since \fy\ has half the period. We expect $(2.140\pm 0.042) \times 10^{-12}$ s s$^{-1}$ and in fact observe $(2.075\pm 0.044) \times 10^{-12}$ s s$^{-1}$, in excellent agreement.

The \fz\ combination frequency is a bit more complicated, given it is a sum of the parent frequencies and not an integer ratio. We expect its rate of change of period with time to be
\begin{equation}
\frac{dP_3}{dt} = \frac{P_3^2}{P_1^2} \frac{dP_1}{dt} + \frac{P_3^2}{P_2^2} \frac{dP_2}{dt} 
\end{equation}
where $P_1 =$ \fa$^{-1}$, $P_2 =$ \fb$^{-1}$, and $P_3 =$ (\fz)$^{-1}$. We thus expect a rate of $(1.027\pm 0.099) \times 10^{-12}$ s s$^{-1}$ and in fact observe $(1.109\pm 0.030) \times 10^{-12}$ s s$^{-1}$. As with \fy, our observed rates match the predictions exactly within the errors.

This lends convincing evidence that these nonlinear combination frequencies are not independent modes themselves, but rather artifacts of some nonlinear distortion in the outer layers of the star. They are thus directly tied to the parent modes, and evolve at exactly the same rate as these parent modes. This is an important result that convincingly demonstrates the nonlinear nature of these signals.

In fact, an extensive theory explaining the creation of nonlinear combination frequencies in DAVs was introduced by \citet{Brickhill92}. He attributed the nonlinearities to the changing thickness of the star's convection zone, which acts to distort a linear input signal. Physically, local surface temperature variations lead to changes in the depth of the convection zone, which both absorb and release energy. Because this process is so sensitive to temperature ($\propto T^{-90}$ for a DAV; \citealt{Montgomery05}), these variations in the depth of the convection zone lead to nonlinear effects in the observed light curve. An alternate explanation that may play a role for the hottest DAVs is given by \citet{Brassard95} and \citet{FontBrass08}. They invoke the nonlinear response of the flux observed in a passband to the temperature (i.e., the ``$T^4$'' nonlinearity). Regardless of the source of these nonlinearities, our result confirms that the phase evolution of these combination frequencies mirrors that of their parent modes. These combination frequencies are thus directly tied to their parent modes.

The presence of nonlinear combination frequencies may also be useful in identifying the spherical degree ($\ell$) and azimuthal order ($m$) of the parent modes. In an investigation of eight DAVs, \citet{Yeates05} found that the amplitudes of combination frequencies are too large to be created by the $T^4$ nonlinearity, and are probably due to the higher-order temperature dependence of the convection zone. They employed the formalism of \citet{Wu01}, who derived an analytical expression for the amplitudes of the combination frequencies in the context of the Brickhill framework of a depth-varying convection zone.  Using the ratio of the amplitudes of the parent and combination frequencies, Yeates et al.\ derived constraints on the $\ell$ and $m$ value of the parent mode.

With the same goal in mind, we have used our own code to automatically search for combinations of $\ell$ and $m$ that best fit the observed amplitudes in \wdo. This code uses a genetic algorithm to efficiently sample parameter space; see \S4.1 of \citet{Provencal12} for more details. We run this algorithm 1000 times so that we obtain 1000 estimates of the best-fit parameters.

\begin{figure}[t]
\centering{\includegraphics[width=0.79\columnwidth,angle=-90]{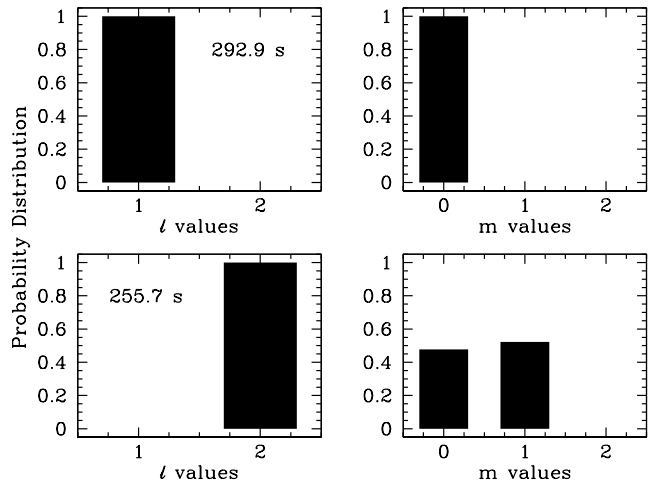}}
\caption{A probability distribution of $\ell$ and $m$ values for the best 7\% of fits using the amplitude ratio of the nonlinear combination frequencies to the parent modes, a method similar to the one used in \citet{Yeates05}. The amplitude ratios for the 292.9 s mode, \fa, are best explained by an $\ell=1$, $m=0$ mode (top two panels). The amplitude ratios for the 255.7 s mode, \fb, indicate this is an $\ell=2$ mode, although there is more ambiguity about the azimuthal order of this mode (bottom panels). \label{fig:modeamps}}
\end{figure}

The method of \citet{Wu01} is only applicable to second-order combinations, i.e., \fy\ or \fz\, but not $3 f_1$.  Even with just three independent combination amplitudes we are able to find constraints. Consistently, we best reproduce the observed amplitude ratios (we have used the amplitudes of the 2009-2012 group, provided in Section~\ref{sec:FT}) when we allow the \fa\ parent mode at 292.9 s to be an $\ell = 1$, $m = 0$ mode.  We also find the \fb\ parent mode at 255.7 s to be an $\ell = 2$, $m = 0, \pm 1$ mode. This is shown in Figure~\ref{fig:modeamps}, where we plot the distribution of the top 7\% of fits with the lowest $\chi^2$, interpreting this as a probability distribution for the identifications.

In addition, we have implemented the convective nonlinear light curve fitting code of \citet{Montgomery05} and \citet{Montgomery10} to generate a synthetic light curve based on the parent modes in \wdo. This method investigates the time-averaged thermal response time of the convection zone, $\tau_c$, which relates to the mass and depth of the convection zone. In principle, it should be superior to the perturbative method of \citet{Wu01} in that it makes use of all higher-order amplitudes and their phases. These nonlinear fits produce estimates for the $\ell$ and $m$ values of the parent modes, the time-averaged convective timescale $\tau_0$, the temperature sensitivity exponent $N$, and the inclination of the pulsation axis to our line of sight $\theta$.

Fitting our light curves from the 2011b subgroup, we find an identical match to our previous method. The 292.9 s mode is best represented by an $\ell = 1$, $m = 0$ mode and the \fb\ parent mode at 255.7 s best fits as an $\ell = 2$, $m = \pm 1$ mode. Additionally, these fits find $\tau_0 = 137.8\pm11.8$ s, $N = -82.0\pm3.6$, and $\theta = 34.6\pm2.5^{\circ}$.

We thus adopt an interpretation that \fa\ is an $m=0$ dipole ($\ell=1$) mode and \fb\ is a quadropole ($\ell=2$) mode. The constraints on the azimuthal order of \fb\ are less conclusive. We will use this information to help constrain our asteroseismic models, which will in turn inform our attempts at understanding the observed $dP/dt$.

\subsection{Expected Rates of Period Change}
\label{sec:expectedPdot}

We can immediately rule out the possibility that an unseen external body is vigorously tugging \wdo\ away from our line of sight, causing a light-travel time effect that is manifest as the increasing delay in pulse arrival times: Rapid space motion would affect all modes identically, and the periodic modulation in the mode \fb\ is more than an order of magnitude slower than that of \fa, as we found in Section~\ref{sec:rates}. We must be observing a phenomenon internal to the WD.

We expect cooling to be the dominant evolutionary effect of an isolated DAV, and this cooling timescale should have a direct effect on the period evolution of modes in a pulsating WD. To first order, as a WD cools it becomes more degenerate. This causes a decrease in the square of the \bruntv\ frequency, $N^2$. The \bruntv\ frequency is essentially the oscillatory frequency of an adiabatic fluid displacement in a convectively stable medium. In the limit of low temperature and complete degeneracy, where pressure does not depend on temperature, a displaced fluid element would remain in pressure and density equilibrium with its surroundings. It would stay put and thus have zero oscillatory motion --- and thus a zero $N^2$. Thus, $N^2$ decreases with increasing degeneracy, and therefore decreasing temperature (see \citealt{Kim07}).

The period of a $g$-mode pulsation is inversely proportional to the \bruntv\ frequency \citep{Unno89}. Therefore, cooling will cause a secular increase in a pulsation period with time, which can be estimated from the cooling timescale predicted from Mestel theory, as well as full evolutionary models. Predicted rates of period change for DAVs undergoing simple cooling are of order 10$^{-15}$ s s$^{-1}$; we find observational evidence to support this timescale from observations of G117-B15A and R548 \citep{Kepler11,Mukadam12}. The 215.2 s mode in G117-B15A and the 213.1 s mode in R548 are most likely either an $\ell=1, k=1$ mode or a $\ell=1, k=2$ mode, depending on the adopted hydrogen layer mass \citep{Kim08,Romero12}.

We note that the work of monitoring long-term phase evolution of DAVs requires a mode coherent in amplitude and phase. This is usually not the case for the cooler DAVs, which evidence strong amplitude modulation and a corresponding lack of stability in phase (e.g. \citealt{Kleinman98,Dolez06}). One cool DAV has, for at least a brief time, shown enough phase coherence to create a meaningful \omc\ diagram: The 615.2 s mode in G29-38 \citep{Kleinman95}, which showed a coherent but extremely rapid phase change, more than 200 s over less than three months of monitoring in 1988 \citep{Winget90}. There is also evidence that the 274.8 s mode in R548 has a $dP/dt > 10^{-13}$ s s$^{-1}$ \citep{Mukadam03}, and that the 270.5 s and 304.1 modes in G117-B15A have $dP/dt > 3 \times 10^{-14}$ s s$^{-1}$ \citep{Kepler05}. Still, most modes in DAVs observed to date simply do not show enough coherence to create stable \omc\ diagrams.

WD stars with heavier cores should cool faster than those that are lighter. Indeed, spectral fits show that \wdo, at $\simeq 0.71$ \msun, is slightly more massive than G117-B15A and R548, which are between 0.59\,--\,0.60 \msun. Still, even a 1.4 \msun\ Fe-core WD would have a $dP/dt$ from cooling of less than $9 \times 10^{-15}$ s s$^{-1}$ if the WD has \teff\ $=11800$ K \citep{Bradley92}.

Our observed rate of change of period with time for \fa\ in \wdo\ exceeds 10$^{-12}$ s s$^{-1}$, more than two orders of magnitude faster than predicted from cooling alone. It is also more than an order of magnitude faster than expected from an avoided crossing \citep{BradleyWinget91} or by invoking excess cooling from exotic particles such as axions \citep{Kim08,Corsico12}. We thus interpret this discovery as evidence of an additional physical timescale acting on the evolution of the pulsation periods in \wdo.

\subsection{Asteroseismology of WD~0111+0018}

Both the 292.9 s and 255.7 s modes of \wdo\ likely probe different regions of the star, so mode identification (specifically the radial order, $k$) would yield further insight into understanding these anomalous rates of period change. Unfortunately, asteroseismology is made more difficult by the presence of just two independent modes; there are far more free parameters than observations.

Still, we have attempted to match the observed periods to adiabatic pulsation models with some assumptions and the constraints provided by our spectroscopic mass and temperature determinations. Our two best matches to the observed periods have the following properties:

Solution 1: \fa\ is an $\ell=1$, $k=5$ mode ($m=0$ occurs at 293.4 s), while \fb\ is an $\ell=2$, $k=8$ mode ($m=0$ occurs at 255.0 s). This WD model has a stellar mass of 0.710 \msun, an effective temperature $T_{\rm  eff} = 11630$  K, a He envelope mass of $M_{\rm He}/M_* = 10^{-2.1}$, and a H envelope mass of $M_{\rm H}/M_* = 10^{-4.5}$.

Solution 2: \fa\ is an $\ell=2$, $k=10$ mode ($m=0$ occurs at 293.7 s), while \fb\ is an $\ell=2$, $k=8$ mode ($m=0$ occurs at 255.0 s). This WD model has a stellar mass of 0.720 \msun, an effective temperature $T_{\rm  eff} = 11610$  K, a He envelope mass of $M_{\rm He}/M_* = 10^{-2.0}$, and a H envelope mass of $M_{\rm H}/M_* = 10^{-4.5}$.

Additionally, we can use our spectroscopic constraints to explore the models of \citet{Romero12} of a 0.705 \msun, $11810$ K WD with a thick ($10^{-4.445}$ \mstar) hydrogen layer mass. Similarly, we find that the 292.9 s mode is either an $\ell=2, k=10$ or $\ell=1, k=5$ mode. The 255.7 s mode is either an $\ell=2, k=8$ or $\ell=1, k=4$ mode. We caution that because these solutions are consistent does not imply they are correct --- our assumptions may not indeed be valid. But these solutions will help inform our analysis of what could be causing the rapid period change in \wdo.

\begin{figure}[t]
\centering{\includegraphics[width=\columnwidth]{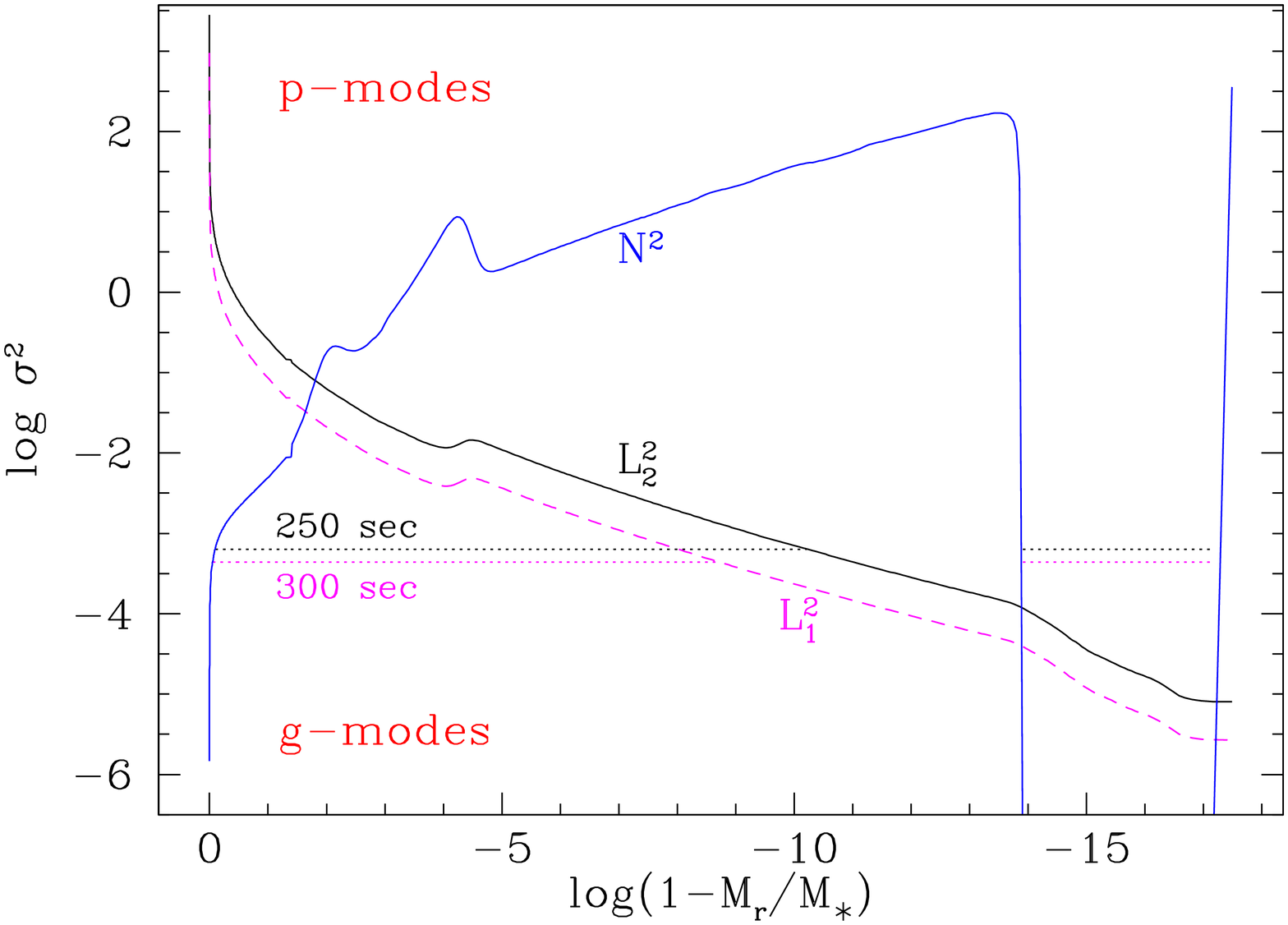}}
\caption{A propagation diagram for a representative asteroseismic fit, a WD model with stellar mass 0.710 \msun, an effective temperature $T_{\rm  eff} = 11630$ K, a He envelope mass of $M_{\rm He}/M_* = 10^{-2.1}$, and a H envelope mass of $M_{\rm H}/M_* = 10^{-4.5}$. The run of the \bruntv\ frequency is shown in blue. The run of the Lamb (acoustic) frequency for an $\ell=1$ mode is shown as a dashed magenta line, while the Lamb frequency for an $\ell=2$ mode is shown as a solid black line. The horizontal axis is in fractional mass units; the surface is to the far right, while the center is to the far left. The base of the convection zone can be seen as the sharp drop in $N^2$ around $(1 - M_{\rm r}/M_*) \simeq 10^{-14}$. \label{fig:propdiag}}
\end{figure}

Using our spherical degree ($\ell$) identifications described in Section~\ref{sec:combs}, we can break the degeneracy in \fa\ and suggest that the 292.9 s periodicity represents an $\ell=1, k=5$ mode, while the the 255.7 s periodicity is an $\ell=2, k=8$ mode. In order to visualize how each such mode samples the star, we have included a propagation diagram in Figure~\ref{fig:propdiag}. We have marked where a $\sim$250 s and a $\sim$300 s $g$-mode for different $\ell$ values would probe the star (where $\sigma^2 < N^2,L_{\ell}^2$).

\begin{figure}[t]
\centering{\includegraphics[width=0.8\columnwidth]{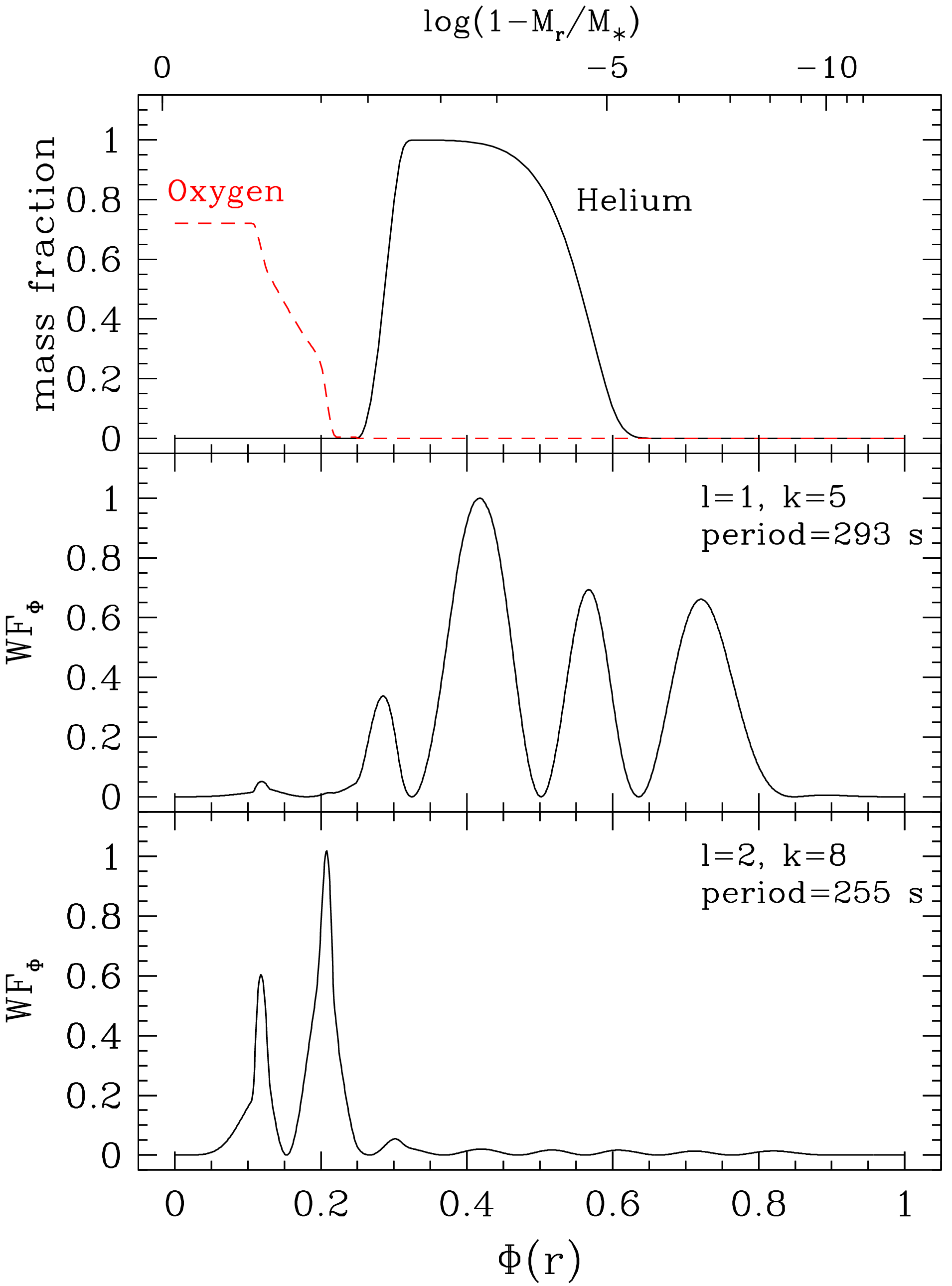}}
\caption{We can investigate how each pulsation samples the star by plotting a weight function (${\rm WF_{\Phi}}$) for each mode. The top panel shows the chemical transition zones as a function of depth in the star. The bottom two panels show the weight functions for our best models (which are only marginally constrained) to represent \fa\ and \fb, respectively. All plots are shown as a function of the ``normalized buoyancy radius,'' $\Phi({\rm  r})$, described in \citet{MikeMon03}, where 0 is the center and 1 is the surface of the WD. The corresponding mass fraction is shown at the top of this figure. \label{fig:wtfxn}}
\end{figure}

However, an even more informative display of how each mode could probe the star differently can be found in Figure~\ref{fig:wtfxn}. Here we plot, as a function of the ``normalized buoyancy radius'' (see \citealt{MikeMon03}), the weight functions for $N^2$ (see Equation 8c of \citealt{KWH85}) for our best identifications of the modes in \wdo. The weight function essentially shows where in the star an excited pulsation is most resonant; it shows the sensitivity of a mode's period to changes in the equilibrium model.

Immediately evident from this exercise is that the majority of the energy of the 292.9 s mode, \fa, is located in the He layer or the region of the He-H transition. Conversely, nearly all the energy of the 255.7 s mode, \fb, is resonant much deeper in the star, inside the C/O-He transition. This suggests that the 255.7 s mode energy is predominantly resonant inside the core, while the 292.9 s mode samples significantly more of the surface layers of the star. The 255.7 s mode would therefore be much less sensitive to changes in the outer regions of the WD.

\subsection{Additional Mechanisms for Rapid Period Change}

Given the insensitivity of the 255.7 s mode to changes in the outer layer of the star, one possibility for the new timescale represented by the fast rate of change of period with time in \wdo\ involves a change in the surface layers of the star at a rate much faster than from cooling alone but much slower than the timescales associated with the pulsations themselves. For instance, the depth of the convection zone of the WD could change gradually, perhaps as a result of a change in the global magnetic field of the WD. We have seen evidence for rapid changes in the pulsation profile of WDs in the past: especially notable is the so-called ``sforzando'' event seen in GD 358 in 1996, in which this DBV changed from a highly multi-periodic pulsation spectrum to a higher-amplitude, monoperiodic pulse shape, indicating a rapid thinning of the convection zone during this event \citep{Provencal09,Montgomery10}.

We have investigated the plausibility of this scenario by comparing two slightly different theoretical models. Both model WDs have identical masses (0.71 \msun), effective temperatures (11810 K), and chemical profiles. However, we have changed the efficiency of convection by varying the mixing length parameter: in our initial model we use ML2/$\alpha$=1.23 and in the other we use ML2/$\alpha$=1.12. This effectively, albeit superficially, causes the second model to have a slightly thinner convection zone.

We take the difference of the periods computed by the first model with the periods from the second to estimate the effect these changes would have on different modes. We find that even the shortest period $\ell=1, k=1$ pulsations would change in mode period by more than 1 ms from one model to the next. This result can reproduce the observed rates of period change in \wdo\ if the timescale of this change in the depth of the convection zone acts on the order a few decades.

One way to perturb the overall depth of the convection zone is to introduce a gradual change in the global magnetic field strength. In fact, increasing the global magnetic field from 0 G to 1 kG would effectively change the depth of the convection zone by the same amount as changing the convective efficiency from ML2/$\alpha$=1.23 to ML2/$\alpha$=1.12 (Montgomery \& Vishniac, in preparation). It is possible then that the rapid period change we observe in \wdo\ could be related to a secular (or long-term periodic) change in the strength of the magnetic field in the star. If this is indeed a byproduct of a stellar cycle, we might expect our heretofore secular increase to at some point turn over or in fact become discontinuous when the magnetic cycle changes course. Future observations of \wdo\ will help constrain this hypothesis.

Another mechanism that could affect one mode differently than another, in the way we see in \wdo, is a changing rotation rate. Rotation will act to break spherical symmetry, splitting non-radial pulsations into ($2\ell + 1$) azimuthal components that would normally be degenerate with the $m=0$ component \citep{Hansen77}. However, if the rotation rate were somehow to change (a spin-up or spin-down), only the $m \ne 0$ components would be affected. If \fa\ is an $m=\pm1$ mode and \fb\ is an $m=0$ mode, a changing rotation rate could explain the large difference in rates of period change between the two.

Our observed rate of change of period with time for \fa\ would require the rotation of the WD to change at a rate $|\dot{P}$$_{\rm rot}| > 0.3$ s yr$^{-1}$. Such a rapid change would require a tremendous reservoir of angular momentum to redistribute, and some place to deposit this energy. This appears especially unlikely given the observational result that the hot pulsating white dwarf PG1159-035 rotates as a solid body through more than 97.5\% of its mass, indicating that WDs have very little angular momentum, even at \teff\ $> 80,000$ K \citep{Charpinet09}. 

Unfortunately, we do not observe any other multiplet components for \fa\ or \fb, making it difficult to definitively rule out this rotational effect\footnote{Due to inclination effects, it is not uncommon to see an $m=\pm1$ component of a rotationally split multiplet and not the $m=0$ central component. This is well illustrated for different $\ell$,$m$ values in Figure 1 of \citet{Brassard95}. It is also possible that only one component of a multiplet is excited.}. But our results from the convective light curve fitting discussed in Section~\ref{sec:combs} did yield predictions for the $m$ value of each mode: $m = 0$ for \fa\ and $m = 0, \pm 1$ for \fb. This contradicts the values needed to invoke a changing rotation rate to explain the different rates of period change between these modes.

Some additional effects intrinsic to the star that could cause phase variations include linear interference (such as beating of closely spaced modes), nonlinear interactions with other modes (such as mode coupling) and a precession of the rotation axis with our line of sight \citep{Winget90}. Each of these cases would simultaneously affect the amplitude of a given mode. Since we do not see evidence of the amplitudes evolving identically with the phase (comparing Figure~\ref{fig:omc} with Figure~\ref{fig:amps}), none of these are likely responsible for the inferred high rates of period change in \wdo.



\section{Conclusions}

As part of a search for substellar companions to isolated WDs, we have discovered a DAV, \wdo, with an unexpectedly fast rate of change of period with time. We see this anomalously high rate reflected in not only the highest-amplitude mode in the star at 292.9 s but also the two nonlinear combination frequencies that sample this parent mode. All three periodicities are changing at a rate faster than 10$^{-12}$ s s$^{-1}$, more than two orders of magnitude more quickly than expected from cooling alone in a DAV.

The pulsation period of the other parent mode at 255.7 s is increasing at a rate at least an order of magnitude more slowly, which allows us to rule out an extrinsic source of this rapid period change. Instead we are likely observing a physical phenomena intrinsic to the star acting on a timescale we have not yet observed in a DAV.

This discovery indicates that our understanding of the expected rates of period change for the slow and simply evolving DAVs is far from complete. It also complicates, in a macroscopic sense, our ability to infer the contributions of exotic weakly interacting particles such as axions to anomalies in the rate of change of period with time of DAVs.

We have attempted to use the nonlinear combination frequencies present in this star as probes of the spherical degree of the two parent modes. Using convective light curve fitting \citep{Montgomery05}, we find that the 292.9 s mode is best represented as an $\ell = 1$, $m = 0$ mode and that the \fb\ parent mode at 255.7 s best fits as an $\ell = 2$, $m = \pm 1$ mode. This identification, if confirmed, would rule out a changing rotation rate as the explanation of the rapid period change in the 292.9 s mode.

Guided by the spherical degrees we identify by this convective light curve fitting we have found a best-fit asteroseismic model, which is admittedly poorly constrained. Still, this exercise suggests that \fa\ is an $\ell=1,k=5$ mode and \fb\ is an $\ell=2,k=8$ mode. We construct a weight function, and show that this solution for \fb\ is almost entirely resonant inside the core, and is thus much less sensitive than \fa\ to changes in the outer regions of the WD.

We discuss a scenario where a gradual thinning of the convection zone could be responsible for the rapid rates of period change observed in this DAV. This thinning could be caused by increasing the global magnetic field by less than 1 kG over a decade. Such a hypothesis is hard to test empirically, although if this magnetic field change is related to a stellar magnetic activity cycle, we would expect an eventual turnover or discontinuity in the observed \omc\ diagram.

Independent of the cause of this unexpected period change, we have for the first time shown that the evolution of nonlinear combination frequencies in a DAV matches the evolution of their parent modes. This behavior confirms that these combination frequencies are not independent pulsation modes but rather artifacts from some nonlinear distortion occurring in the star.


\acknowledgments

We are especially grateful to all those whose time in West Texas made this result possible: S. E. Thompson, C. M. Yeates, K. I. Winget, Davis Winget, R. E. Nather, Elizabeth J. Jeffery, Ross E. Falcon, and G. F. Miller. We thank E. L. Robinson, Anjum S. Mukadam, Dennis Sullivan, and Ross E. Falcon for helpful discussions, and acknowledge the McDonald Observatory staff for their tireless support, especially Dave Doss and John Kuehne. This work is supported by the Norman Hackerman Advanced Research Program, under grants 003658-0255-2007 and 003658-0252-2009, by a grant from the NASA Origins Program, NAG5-13094, and by the National Science Foundation, under grant AST-0909107.


\begin{thebibliography}{}

\bibitem[Althaus et 
al.(2010)]{Althaus10} Althaus, L.~G., C{\'o}rsico, A.~H., Isern, J., \& Garc{\'{\i}}a-Berro, E.\ 2010, \aapr, 18, 471 

\bibitem[Bergeron et al.(2004)]{Bergeron04} Bergeron, P., Fontaine, G., Bill{\`e}res, M., Boudreault, S., \& Green, E.~M.\ 2004, \apj, 600, 404 

\bibitem[Bischoff-Kim et al.(2008)]{Kim08} Bischoff-Kim, A., 
Montgomery, M.~H., \& Winget, D.~E.\ 2008, \apj, 675, 1505 

\bibitem[Bradley 
\& Winget(1991)]{BradleyWinget91} Bradley, P.~A., \& Winget, D.~E.\ 1991, \apjs, 75, 463 

\bibitem[Bradley et al.(1992)]{Bradley92} Bradley, P.~A., Winget, D.~E., \& Wood, M.~A.\ 1992, \apjl, 391, L33 

\bibitem[Brassard et al.(1995)]{Brassard95} Brassard, P., Fontaine, G., \& Wesemael, F.\ 1995, \apjs, 96, 545 

\bibitem[Brickhill(1992)]{Brickhill92} Brickhill, A.~J.\ 1992, \mnras, 259, 529

\bibitem[Charpinet et al.(2009)]{Charpinet09} Charpinet, S., 
Fontaine, G., \& Brassard, P.\ 2009, \nat, 461, 501 

\bibitem[C{\'o}rsico et al.(2012)]{Corsico12} C{\'o}rsico, A.~H., 
Althaus, L.~G., Miller Bertolami, M.~M., et al.\ 2012, \mnras, 424, 2792 

\bibitem[Dolez et al.(2006)]{Dolez06} Dolez, N., Vauclair, G., Kleinman, S.~J., et al.\ 2006, \aap, 446, 237 

\bibitem[Fontaine et al.(1985)]{Fontaine85} Fontaine, G.,
Bergeron, P., Lacombe, P., Lamontagne, R., \& Talon, A.\ 1985, \aj, 90,
1094

\bibitem[Fontaine 
\& Brassard(2008)]{FontBrass08} Fontaine, G., \& Brassard, P.\ 2008, \pasp, 120, 1043 

\bibitem[Gianninas et al.(2005)]{Gianninas05} Gianninas, A.,
Bergeron, P., \& Fontaine, G.\ 2005, \apj, 631, 1100

\bibitem[Gianninas et al.(2006)]{Gianninas06} Gianninas, A.,
Bergeron, P., \& Fontaine, G.\ 2006, AJ, 132, 831 

\bibitem[Hansen et al.(1977)]{Hansen77} Hansen, C.~J., Cox, J.~P., \& van Horn, H.~M.\ 1977, \apj, 217, 151

\bibitem[Jones et al.(1989)]{Jones89} Jones, P.~W., Hansen, C.~J., Pesnell, W.~D., \& Kawaler, S.~D.\ 1989, \apj, 336, 403

\bibitem[Kanaan et 
al.(2002)]{Kanaan02} Kanaan, A., Kepler, S.~O., \& Winget, D.~E.\ 2002, \aap, 389, 896 

\bibitem[Kawaler et al.(1985)]{KWH85} Kawaler, S.~D., Winget, 
D.~E., \& Hansen, C.~J.\ 1985, \apj, 295, 547 

\bibitem[Kepler et al.(1991)]{Kepler91} Kepler, S.~O., et al.\ 1991, \apjl, 378, L45 

\bibitem[Kepler et al.(2005)]{Kepler05} Kepler, S.~O., et al.\ 2005, \apj, 634, 1311

\bibitem[Kepler et al.(2011)]{Kepler11} Kepler, S.~O., 2011 in Shibahashi H., ed., ASP Conf. Ser. Proceedings of the 61st Fujihara Seminar: Progress in Solar/Stellar Physics with Helio and Asteroseismology. Astron. Soc. Pac., San Francisco, in press

\bibitem[Kim(2007)]{Kim07} Kim, A.\ 2007, PhD thesis, Univ. Texas, Austin

\bibitem[Kleinman(1995)]{Kleinman95} Kleinman, S.~J.\ 1995, PhD thesis, Univ. Texas, Austin

\bibitem[Kleinman et al.(1998)]{Kleinman98} Kleinman, S.~J., 
Nather, R.~E., Winget, D.~E., et al.\ 1998, \apj, 495, 424 

\bibitem[Koester \& Holberg(2001)]{Koester01} Koester, D., \&
Holberg, J.~B.\ 2001, ASP Conf.~Ser.~226: 12th European Workshop on White
Dwarfs, 226, 299 

\bibitem[Lenz 
\& Breger(2005)]{Lenz05} Lenz, P., \& Breger, M.\ 2005, Communications in Asteroseismology, 146, 53

\bibitem[Mestel(1952)]{Mestel52} Mestel, L.\ 1952, \mnras, 112, 583 

\bibitem[Montgomery et al.(2003)]{MikeMon03} Montgomery, M.~H., 
Metcalfe, T.~S., \& Winget, D.~E.\ 2003, \mnras, 344, 657 

\bibitem[Montgomery(2005)]{Montgomery05} Montgomery, M.~H.\ 2005, 
\apj, 633, 1142 

\bibitem[Montgomery et al.(2010)]{Montgomery10} Montgomery, M.~H., et al.\ 2010, \apj, 716, 84 

\bibitem[Mukadam et al.(2003)]{Mukadam03} Mukadam, A.~S., Kepler, 
S.~O., Winget, D.~E., et al.\ 2003, \apj, 594, 961

\bibitem[Mukadam et al.(2004)]{Mukadam04} Mukadam, A.~S., Winget,
D.~E., von Hippel, et al.\ 2004, \apj, 612, 1052 

\bibitem[Mukadam et al.(2006)]{Mukadam06} Mukadam, A.~S., Montgomery, M.~H., Winget, D.~E., Kepler, S.~O., \& Clemens, J.~C.\ 2006, \apj, 640, 956

\bibitem[Mukadam et al.(2012)]{Mukadam12} Mukadam, A.~S., et al.\ 2012, Journal of Physics Conference Series, in press 

\bibitem[Mullally et al.(2005)]{Mullally05} Mullally, F., 
Thompson, S.~E., Castanheira, B.~G., et al.\ 2005, \apj, 625, 966 

\bibitem[Mullally et al.(2008)]{Mullally08} Mullally, F., Winget, D.~E., De Gennaro, S., Jeffery, E., Thompson, S.~E., Chandler, D., \& Kepler, S.~O.\ 2008, \apj, 676, 573 

\bibitem[Nather \& Mukadam(2004)]{Nather04} Nather, R.~E.~\& Mukadam, A.~S.\ 2004, \apj, 605, 846

\bibitem[Provencal et al.(2009)]{Provencal09} Provencal, J.~L., 
Montgomery, M.~H., Kanaan, A., et al.\ 2009, \apj, 693, 564 

\bibitem[Provencal et al.(2012)]{Provencal12} Provencal, J.~L., 
Montgomery, M.~H., Kanaan, A., et al.\ 2012, \apj, 751, 91 

\bibitem[Robinson(1979)]{Robinson79} Robinson, E.~L.\ 1979, IAU
Colloq.~53: White Dwarfs and Variable Degenerate Stars, 343

\bibitem[Romero et al.(2012)]{Romero12} Romero, A.~D., 
C{\'o}rsico, A.~H., Althaus, L.~G., et al.\ 2012, \mnras, 420, 1462 

\bibitem[Stumpff(1980)]{Stumpff80} Stumpff, P.\ 1980, \aaps, 41, 1 

\bibitem[Thompson \& Mullally(2009)]{Thompson09} Thompson, S.~E., \& Mullally, F.\ 2009, Journal of Physics Conference Series, 172, 012081

\bibitem[Tremblay et al.(2011)]{Tremblay11} Tremblay, P.-E., 
Bergeron, P., \& Gianninas, A.\ 2011, \apj, 730, 128 

\bibitem[Unno et al.(1989)]{Unno89} Unno, W., Osaki, Y., Ando, H., Saio, H., \& Shibahashi, H.\ 1989, Nonradial oscillations of stars, Tokyo: University of Tokyo Press, 1989, 2nd ed., 

\bibitem[Winget et al.(1990)]{Winget90} Winget, D.~E., Nather, 
R.~E., Clemens, J.~C., et al.\ 1990, \apj, 357, 630  

\bibitem[Winget et al.(2003)]{Winget03} Winget, D.~E., et al.\ 2003, Scientific Frontiers in Research on Extrasolar Planets, 294, 59 

\bibitem[Winget et al.(2004)]{Winget04} Winget, D.~E., Sullivan, 
D.~J., Metcalfe, T.~S., Kawaler, S.~D., 
\& Montgomery, M.~H.\ 2004, \apjl, 602, L109 

\bibitem[Winget 
\& Kepler(2008)]{WinKep08} Winget, D.~E., \& Kepler, S.~O.\ 2008, \araa, 46, 157 

\bibitem[Wu(2001)]{Wu01} Wu, Y.\ 2001, \mnras, 323, 248 

\bibitem[Yeates et al.(2005)]{Yeates05} Yeates, C.~M., Clemens, J.~C., Thompson, S.~E., \& Mullally, F.\ 2005, \apj, 635, 1239 

\end{thebibliography}
\end{document}